\documentclass[aps,twocolumn,prl,superscriptaddress,amsmath,amssymb]{revtex4-1}
\def\etal{{\it et\ al.}}

\newcommand{\means}[1]{\langle#1\rangle}

\def\journal #1#2#3#4{\textit{#1} \textbf{#2}, #3 (#4)}
\def\etal{\textit{et~al.}\ }
\def\PR{Phys.\ Rev.}

\def\PRB{Phys.\ Rev.\ B}
\def\PRL{Phys.\ Rev.\ Lett.}

\usepackage{graphicx}
\usepackage{dcolumn}
\usepackage{bm}
\usepackage{amsmath}
\usepackage[usenames]{color}

\usepackage{txfonts}
\usepackage[T1]{fontenc}
\usepackage{xspace}

\usepackage{hyperref}

\usepackage{ulem}

\begin{document}
\title{
Fermionic response from fractionalization in an insulating two-dimensional magnet
}
\author{J. Nasu}
\affiliation{
  Department of Physics, Tokyo Institute of Technology, 
  Meguro, Tokyo 152-8551, Japan
} 
\author{J. Knolle}
\affiliation{
  Department of Physics, Cavendish Laboratory,
  JJ Thomson Avenue, Cambridge CB3 0HE, U.K.
} 
\author{D. L. Kovrizhin}
\affiliation{
  Department of Physics, Cavendish Laboratory,
  JJ Thomson Avenue, Cambridge CB3 0HE, U.K.
} 
\affiliation{
  RRC Kurchatov Institute, 1 Kurchatov Square, Moscow 123182, Russia
} 
\author{Y. Motome}
\affiliation{
  Department of Applied Physics, University of Tokyo, 
  Bunkyo, Tokyo 113-8656, Japan
} 
\author{R. Moessner}
\affiliation{
Max Planck Institute for the Physics of Complex Systems,
 D-01187 Dresden, Germany
 } 
 
\maketitle

\textbf{
Conventionally ordered magnets possess bosonic elementary excitations, called magnons.
By contrast, no magnetic insulators in more than one dimension are known whose excitations are not bosons but fermions.
Theoretically, some quantum spin liquids (QSLs)~\cite{Anderson1973153} --  new topological phases which  can occur when quantum fluctuations preclude an ordered state -- are known to exhibit Majorana fermions~\cite{Kitaev2006} as quasiparticles arising from fractionalization of spins~\cite{Lacroix2011}.
Alas, despite much searching, their experimental observation remains elusive.
 Here, we show that fermionic excitations are remarkably directly evident in experimental Raman scattering data~\cite{PhysRevLett.114.147201} across a broad energy and temperature range in the two-dimensional material $\alpha$-RuCl$_3$.
 This shows the importance of magnetic materials as hosts of Majorana fermions.
 In turn, this first systematic evaluation of the dynamics of a QSL at finite temperature emphasizes the role of excited states for detecting such exotic properties associated with otherwise hard-to-identify topological QSLs.
 }

\begin{figure}[b]
\begin{center}
\includegraphics[width=\columnwidth,clip]{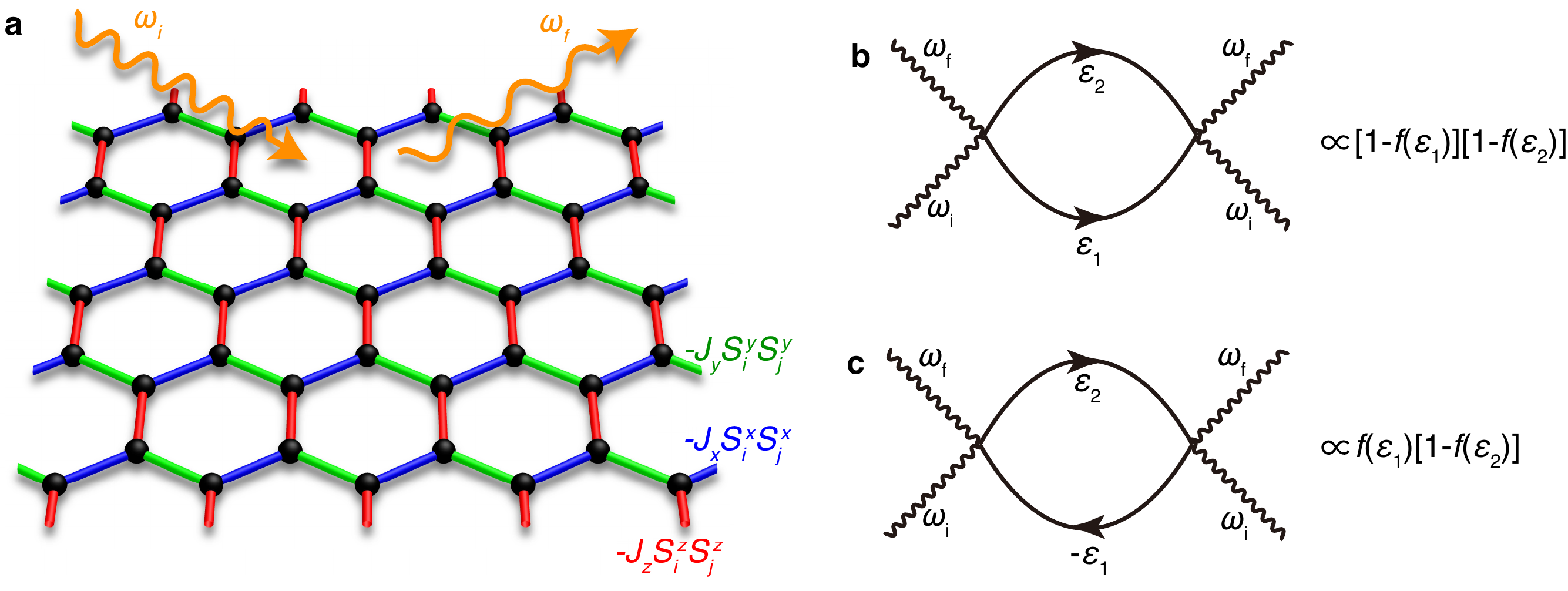}
\end{center}
 \caption{
 \textbf{Schematic pictures for the Kitaev model and Raman processes.}
 \textbf{a}, honeycomb lattice structure.
 Blue, green, and red bonds represent Ising-like interactions between $x$-, $y$-, and $z$-components of the $S=1/2$ spins, respectively.
 Incoming and outgoing photons, whose frequencies are $\omega_i$ and $\omega_f$, are also depicted. 
 \textbf{b} and \textbf{c}, Feynman diagrams of the Raman scattering processes which correspond to a creation or annihilation of a pair of matter fermions [process (A)] and a combination of creation and annihilation of the matter fermions [process (B)], respectively.
 In process (A), a photon scattering creates two fermions with energies $\varepsilon_1$ and $\varepsilon_2$ as shown in \textbf{b}, and hence, the Raman shift $\omega=\omega_f-\omega_i$ is equal to $\varepsilon_1+\varepsilon_2$. 
 In  process (B), the  scattering creates a fermion with energy $\varepsilon_2$ and annihilates a fermion with $\varepsilon_1$ simultaneously as shown in \textbf{c}, and hence, $\omega$ is equal to $\varepsilon_2-\varepsilon_1$.
\label{lattice}
 }
\end{figure}

The Kitaev model has recently attracted attention as a canonical example of a QSL with emergent fractionalized fermionic excitations~\cite{Kitaev2006,PhysRevLett.98.247201}.
The model is defined for $S=1/2$ spins on a honeycomb lattice with anisotropic bond-dependent interactions, as shown in Fig.~\ref{lattice}a~\cite{Kitaev2006}.
Recent theoretical work -- by providing access to properties of excited states -- has predicted signs of Kitaev QSLs in the dynamical response at $T=0$~\cite{PhysRevLett.112.207203,PhysRevLett.113.187201} and in the $T$ dependence of thermodynamic quantities~\cite{PhysRevLett.113.197205,PhysRevB.92.115122}.
However, the dynamical properties at finite $T$ have remained a theoretical challenge as it is necessary to handle quantum and thermal fluctuations simultaneously.
Here, by calculating dynamical correlation functions over a wide temperature range we directly identify signatures of fractionalization in available experimental inelastic light scattering data.

In real materials, Kitaev-type anisotropic interactions may appear through a superexchange process between $j_{\rm eff}=1/2$ localized moments in the presence of strong spin-orbit coupling~\cite{PhysRevLett.102.017205}.
Such a situation is believed to be realised in several materials, such as iridates $A_{2}$IrO$_3$ ($A$=Li, Na)~\cite{PhysRevLett.108.127203,PhysRevLett.109.266406} and a ruthenium compound $\alpha$-RuCl$_3$~\cite{PhysRevB.91.094422,PhysRevLett.114.147201,PhysRevB.90.041112,Banerjee2015}.
These materials show magnetic ordering at a low $T$ ($\sim 10$~K), indicating that some exchange interactions coexist with the Kitaev exchange and give rise to the magnetic order instead of the QSL ground state~\cite{PhysRevLett.105.027204,PhysRevB.84.100406,PhysRevLett.110.097204,PhysRevLett.113.107201}.
Nevertheless, evidence suggests that the Kitaev interaction is predominant (several tens to hundreds of Kelvin)~\cite{PhysRevLett.113.107201,1367-2630-16-1-013056,PhysRevB.88.035107,PhysRevLett.110.097204,PhysRevB.91.241110,Banerjee2015}, which may provide an opportunity to observe the fractional excitations in a quantum paramagnetic state above the transition temperature as a proximity effect of the QSL phase.

\begin{figure}
\begin{center}
\includegraphics[width=\columnwidth,clip]{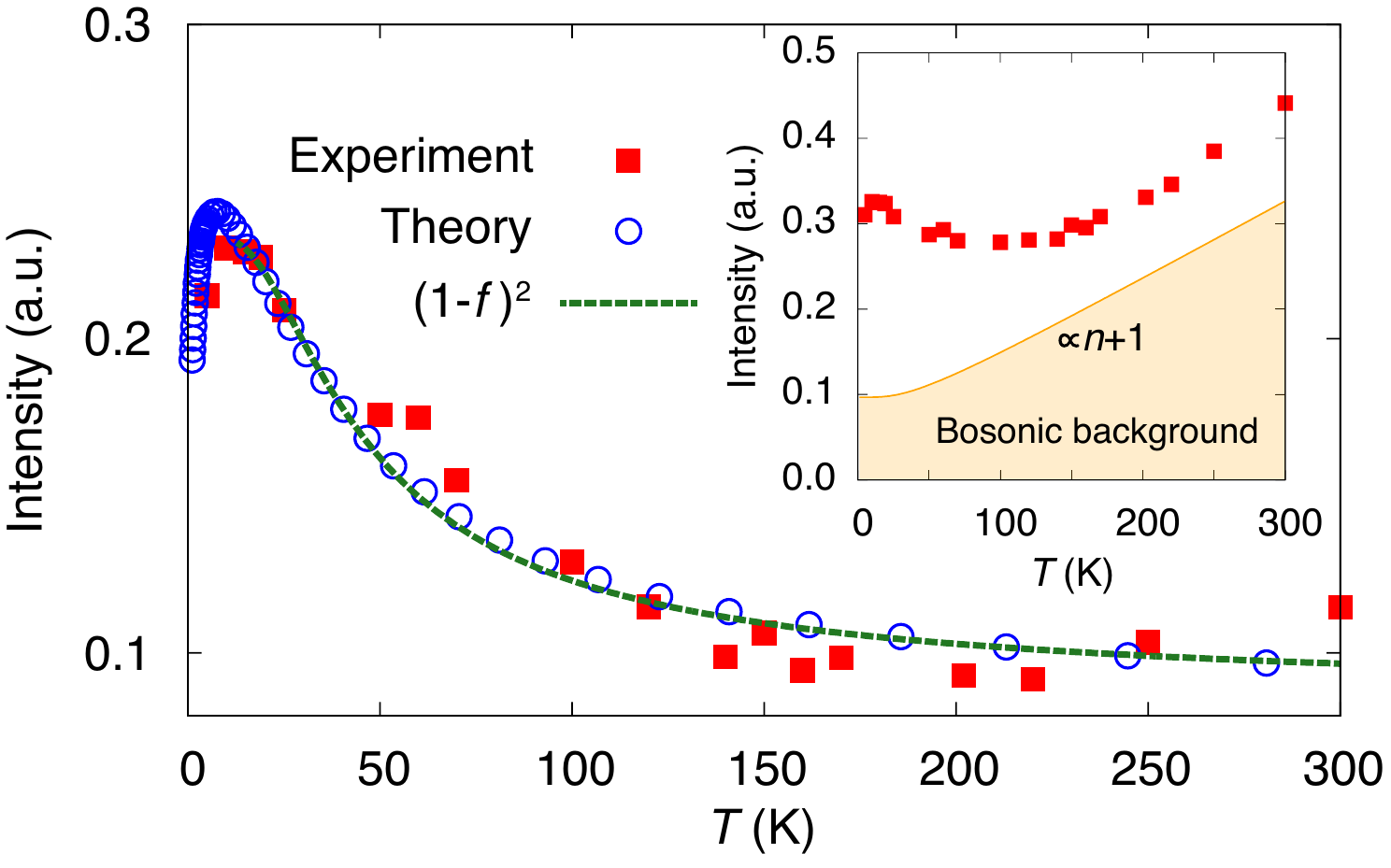}
\end{center}
\caption{
 \textbf{Comparison between the numerical results and the experimental data for $\alpha$-RuCl$_3$.}
 Main panel: blue circles represent  QMC data for a $L=20$ cluster for the integrated Raman intensity $I_{\rm mid}$ shown in Fig.~\ref{phys}\textbf{c}. 
 Red squares are the experimental data in the energy window from 5~meV to 12.5~meV~\cite{PhysRevLett.114.147201}, from which the non-magnetic background is subtracted (see text).
 Green dashed lines represent the fitting by $a_{M}[1-f(\varepsilon_{M}^*)]^2+b_{M}$ (see caption of Fig.~\ref{phys}).
We take $J=10$~meV in calculating $I_{\rm mid}$.
 Inset:  red squares show the experimental raw data and the orange curve indicates the bosonic background.
\label{comp}
 }
\end{figure}

In particular, unconventional excitations were observed by polarized Raman scattering in $\alpha$-RuCl$_3$~\cite{PhysRevLett.114.147201}. 
In this material, N\'eel ordering sets in only at $T_c\sim 14$K, while the Kitaev interaction appears to be much larger than the Heisenberg interaction~\cite{PhysRevB.91.241110,Banerjee2015}, and hence finite-temperature signatures of the Kitaev QSL are expected to be observed in the paramagnetic state persisting in a broad temperature window above $T_c$. 

The inset of Figure~\ref{comp} shows the integrated experimental Raman intensity for $\alpha$-RuCl$_3$ as a function of temperature~\cite{PhysRevLett.114.147201}.
A background contribution, likely due to phonons, has been identified and subtracted~\cite{PhysRevLett.114.147201}, as it persists up to very high $T$ much larger than any magnetic scale.
In this limit, it can be fitted to standard one-particle scattering which is proportional to $n+1$ with $n=1/(e^{\beta\omega}-1)$ being the Bose distribution function. 
The main panel (red symbols) shows the remaining, presumably dominantly magnetic contribution. 

Most remarkably, the $T$ dependence of the spectral weight up to high temperatures (more than an order of magnitude above $T_c$), does not follow the bosonic form expected for conventional insulating magnets in which both magnons and phonons obey Bose statistics. 
It is thus imperative to understand the origin of this anomalous contribution.
This will provide a more direct test of the proximity to QSLs than an asymptotic low-$T$ behaviour which is sensitive to the subdominant 
exchange interactions.

{\bf Results.} 
The main panel of Figure~\ref{comp} provides a comparison of the $T$ dependence of our theoretical results (blue circles) with the experimental data.
The good agreement over a wide temperature range, from just above $T_c$ up to a much higher scale $(\sim  15 T_c)$, offers compelling evidence that our Kitaev QSL theory correctly identifies the nature of fundamental excitations in the form of  fractionalized fermions.
This is further reinforced by noticing that the asymptotic two-fermion-scattering form  $(1-f)^2$, with $f=1/(1+e^{\beta \varepsilon})$ being the Fermi distribution function, is a good fit of the response.
In the following, we outline our calculations and explain how the two-fermion-scattering $T$-dependence emerges as a result of fractionalization.

We investigate the Raman spectrum at finite $T$ for the Kitaev model using quantum Monte Carlo (QMC) simulations which directly utilize the fractionalization of quantum spins into two species of Majorana fermions: itinerant ``matter'' and localized ``flux'' fermions (see Methods for details).
Crucially, the Raman response is elicited only by the itinerant Majorana fermions~\cite{PhysRevLett.113.187201}, which allows us to detect their Fermi statistics more directly than in other dynamical responses~\cite{PhysRevLett.112.207203}.
Below we focus on the case of isotropic exchange couplings, $J_x=J_y=J_z=J$; a small anisotropy plausible in real materials does not alter our main conclusions (see Supplementary information). 
The thermodynamic behaviour exhibits two characteristic crossover $T$-scales originating from fractionalization at $T^{*}/J\sim 0.012$ and $T^{**}/J\sim 0.38$:
the former is related to the condensation of flux Majorana fermions, set by the flux gap $\sim 0.06J$~\cite{Kitaev2006}, while the latter arises from the formation of matter Majorana fermions at much higher $T$, set by their  bandwidth $\sim 1.5J$.

\begin{figure}
\begin{center}
\includegraphics[width=\columnwidth,clip]{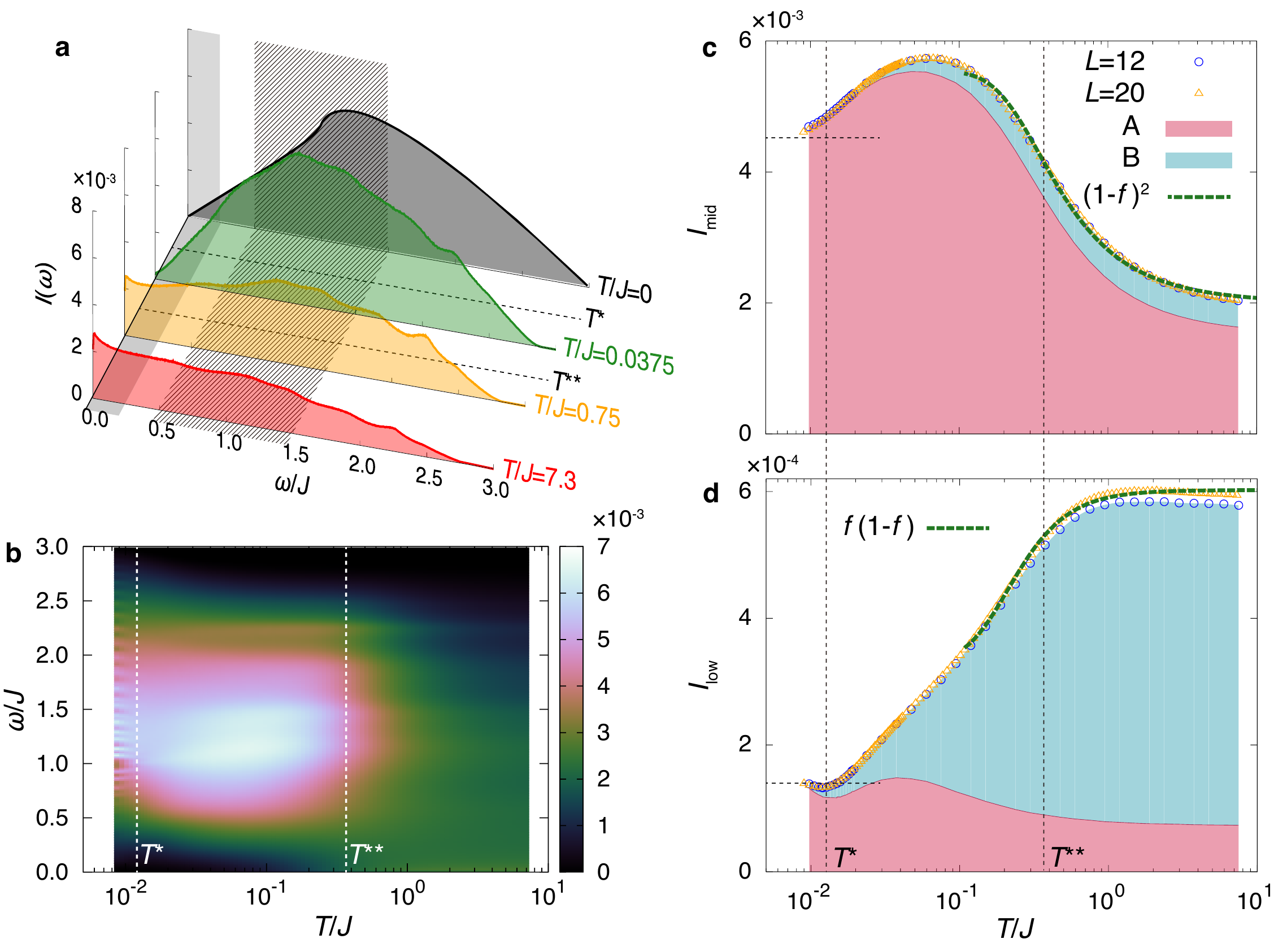}
\end{center}
\caption{
 \textbf{Calculated Raman intensity for finite temperatures.}
 \textbf{a}, $\omega$ dependences of the Raman spectra $I(\omega)$ at several $T$.
 \textbf{b}, contour map of the Raman spectrum $I(\omega)$ in the $T$-$\omega$ plane.
 \textbf{c}, integrated spectral weights $I_{\rm mid}$ for $0.5<\omega/J<1.5$, and \textbf{d}, $I_{\rm low}$ for $0.0<\omega/J<0.25$, whose energy ranges are indicated by the hatched and shaded areas in Fig.~\ref{phys}\textbf{a}, respectively.
 Green dashed lines represent fits with $a_{M}[1-f(\varepsilon_{M}^*)]^2+b_{M}$ for \textbf{c} with $\varepsilon_{M}^*/J=0.62$; and $a_L f(\varepsilon_L^*)[1-f(\varepsilon_L^*)]+b_L$ for \textbf{d} with $\varepsilon_L^*/J=0.42$ (see Supplementary information).
 Here, $f(\varepsilon)=(1+e^{\beta \varepsilon})^{-1}$ is the Fermi distribution function with zero chemical potential.
Horizontal dashed-dotted lines represent the values of $I_{\rm low}$ and $I_{\rm mid}$ at $T=0$~\cite{PhysRevLett.113.187201,PhysRevB.92.094439}.
The red and blue areas in \textbf{c} and \textbf{d} highlight the contributions to the integrated Raman intensities from the processes (A) and (B), respectively.
 Vertical dotted lines indicate two crossover temperatures, $T^*$ and $T^{**}$.
\label{phys}
 }
\end{figure}

Figure~\ref{phys}a shows the QMC data for the Raman spectrum $I(\omega)$ at several $T$.
At $T=0$, it exhibits $\omega$-linear behaviour in the low energy region, due to a linear Dirac dispersion of matter Majorana fermions~\cite{PhysRevLett.113.187201}.
With increasing $T$ above $T^*$, the low energy part increases and the $\omega=0$ contribution becomes nonzero, as shown in the figure for $T/J=0.0375$.
At higher $T$, the broad peak in the intermediate energy range at $\omega/J\sim 1$ is suppressed above $T\sim T^{**}$. 
Indeed, the Raman spectrum at $T/J=0.75$ shows no substantial energy dependence for $0<\omega/J\lesssim 2$, as shown in Fig.~\ref{phys}a.
For higher $T$, the intermediate-to-high energy weight gradually decreases.
The $T$ and $\omega$ dependence of the Raman spectrum is summarized in Fig.~\ref{phys}b. 
The result clearly shows that the broad peak structure is slightly shifted to the low energy side 
above $T^*$ and the spectrum becomes featureless above $T^{**}$.

For further understanding of the $T$ dependence of the Raman spectra, it is helpful to work in a basis of complex matter fermions constructed as a superposition of real Majorana fermions (see Methods).
These elementary excitations determine the $T$-dependence because their occupation (in a fixed background of fluxes) is given by the Fermi distribution function.
In detail, one needs to analyse two different processes contributing to Raman scattering~\cite{PhysRevB.92.094439}:
one consists of creation or annihilation of a pair of fermions [process (A)], with the other a combination of the creation of one fermion and the annihilation of another [process (B)] (see Methods for details).
Process (A) is  proportional to $[1-f(\varepsilon_1)][1-f(\varepsilon_2)]\delta(\omega-\varepsilon_1-\varepsilon_2)$, where $\omega$ is the Raman shift, $\varepsilon_1$ and $\varepsilon_2$ are the energies of fermions (see Fig.~\ref{lattice}b).
Process (B) is proportional to $f(\varepsilon_1)[1-f(\varepsilon_2)]\delta(\omega+\varepsilon_1-\varepsilon_2)$ and vanishes at $T=0$ due to absence of matter fermions in the ground state (see Fig.~\ref{lattice}c).
Because of their different frequency dependence -- e.g., (A) vanishes for $\omega \to 0$ at low $T$ -- their distinct $T$-behaviour can be extracted by looking at different frequency windows.

 Figure~\ref{phys}c shows the $T$ dependence of the integrated spectral weight in the middle energy window, $I_{\rm mid}$ for $0.5<\omega/J<1.5$ (see the hatched region in Fig.~\ref{phys}a).
 The same is used in Fig.~\ref{comp} in accordance with the frequency window for the experimental data with $J=10$~meV.
We emphasize that the value of $J$ is consistent not only with the spectral width and peak position of the Raman continuum at the lowest $T$~\cite{PhysRevLett.114.147201} but also with the inelastic neutron scattering in $\alpha$-RuCl$_3$~\cite{Banerjee2015}.
As shown in Fig.~\ref{phys}c, $I_{\rm mid}$ has a non-monotonic change as a function of $T$: it grows around $T^*$ with increasing $T$, but turns over to decrease above $T/J \sim 0.1$, yielding the shift of the peak structure in $I(\omega)$ to the low energy side shown in Fig.~\ref{phys}b.
We also highlight the contributions from the processes (A) and (B) in Fig.~\ref{phys}c. 
The result clearly indicates that $I_{\rm mid}$ is dominated by the process (A), which supports the scaling with $(1-f)^2$ (see Supplementary information).

Meanwhile, the results presented in Figure~\ref{phys}d covering the low energy window, $I_{\rm low}$ for $0.0<\omega/J<0.25$ (see the shaded region in Fig.~\ref{phys}a), have a different $T$-dependence.
The increase around $T^*$ is because the Dirac semimetallic dip in the itinerant fermion system is filled in due to thermal fluctuations of the flux fermions~\cite{PhysRevB.92.115122}.
Moreover, with increasing $T$, $I_{\rm low}$ saturates around the high-$T$ crossover $T^{**}$.
As shown in Fig.~\ref{phys}d, above $T/J \sim 0.1$ $I_{\rm low}$ is dominated by the process (B), indicating that the $T$ dependence is well fitted by $f(1-f)$.
However, the intensity $I_{\rm low}$, is one order of magnitude smaller than $I_{\rm mid}$.

{\bf Discussion.}
The striking $T$ dependence of the Raman intensity observed in experiments can be naturally attributed to the response from fractionalized fermionic Majorana excitations, dominantly from pairs of creation and annihilation of matter fermions. 
 The $T$ dependence is qualitatively different from that of conventional insulating magnets which show bosonic Raman spectra from two-magnon scattering~\cite{PhysRevB.57.8478}. 
 It is important to note that here we are dealing with a two-dimensional magnet ~\cite{PhysRevB.91.094422,PhysRevB.90.041112,Banerjee2015}.
 In one dimension, there is no such crisp distinction between Bose and Fermi statistics, as in the absence of true exchange processes, bosons with hardcore repulsion are rather similar to fermions obeying the Pauli principle; and on the other hand the roles of topology and order in two dimensions are quite distinct from a one-dimensional case~\cite{WenBook}.

The crucial observation here is that the unexpected fermionic contribution is clearly observed over a remarkably wide $T$ range, more than an order of magnitude higher than the transition temperature into the incidental low-temperature N\'eel order. 
This approach is distinct from the conventional quest for exotic properties of QSLs, where the experimental hallmark of fermionic excitations has mainly been pursued in asymptotic $T$ behaviour, e.g., in the $T$-linear specific heat for temperatures much lower than the interaction energy.
However, the low-$T$ analyses of such thermodynamic quantities are further complicated by the need to distinguish between QSLs, glassy behaviour, spurious order, and other low energy contributions typified, e.g., by nuclear spins.
Our finding provides a direct way of identifying QSL behaviour, and in particular, the presence of fermionic excitations.
This, we hope, will stimulate further studies of other dynamical quantities in the wide $T$ range~\cite{Banerjee2015} as well as studies of other candidate materials like $A_2$IrO$_3$ ($A$=Li, Na)~\cite{PhysRevB.87.220407}.

\textbf{Methods.}
The Hamiltonian of the Kitaev model on the honeycomb lattice is given by
\begin{align}
{\cal H}=-J_x\sum_{\means{jk}_x}S_j^x S_k^x-J_y\sum_{\means{jk}_y}S_j^y S_k^y-J_z\sum_{\means{jk}_z}S_j^z S_k^z,
\label{eq:H_spin}
\end{align}
where $\bm{S}_j =(S_j^x, S_j^y, S_j^z)$ represents an $S=1/2$ spin on site $j$, and $\means{jk}_\gamma$ stands for a nearest-neighbour $\gamma(=x,y,z)$ bond shown in Fig.~\ref{lattice}a~\cite{Kitaev2006}.
By using the Jordan-Wigner transformation and introducing two kinds of Majorana fermions $c_j$ and $\bar{c}_j$~\cite{PhysRevB.76.193101,PhysRevLett.98.087204}, the model is rewritten as
 \begin{align}
  {\cal H}=\frac{iJ_x}{4}\sum_{(jk)_x}c_j c_k-\frac{iJ_y}{4}\sum_{(jk)_y}c_j c_k-\frac{iJ_z}{4}\sum_{(jk)_z} \eta_r c_j c_k,
  \label{eq:H_Majorana}
 \end{align}
 where $(jk)_\gamma$ is the nearest-neighbour pair satisfying $j<k$ on the $\gamma$ bond, and $\eta_r=i\bar{c}_j\bar{c}_k$ is a $Z_2$ variable defined on the $z$ bond ($r$ is the label for the bond), which takes $\pm 1$.
Eq.~(\ref{eq:H_Majorana}) describes free itinerant Majorana fermions coupled to classical $Z_2$ variables $\{\eta_r\}$. 
While the configurations of $\{\eta_r\}$ are thermally disturbed away from the ground state configuration with all $\eta_r=1$, the thermodynamic behaviour can be obtained by properly sampling $\{\eta_r\}$ as follows.
As the Hamiltonian for a given configuration of $\{\eta_r\}$ is bilinear in terms of $c$ operators, it is easily diagonalized as
   \begin{align}
 {\cal H}(\{\eta_r\})=\sum_\lambda \varepsilon_\lambda\left( f_\lambda^\dagger f_\lambda -\frac{1}{2}\right).
 \end{align}
Here, we introduce complex matter fermions $f_\lambda$ with the eigenenergies $\varepsilon_\lambda(\geq 0)$, which are related to $c$ by
\begin{align}
 c_j=\sum_\lambda \left(X_{j\lambda} f_\lambda + X_{j\lambda}^* f_\lambda^\dagger\right),\label{eq:1}
\end{align}
where $X_{j\lambda}$ is introduced so as to diagonalize the Hamiltonian.
Then, we evaluate the free energy $F_f(\{\eta_r\})=-\beta^{-1}\ln Z_f(\{\eta_r\})$ for the configuration $\{\eta_r\}$, where $Z_f(\{\eta_r\})={\rm Tr}_{\{c_j\}}e^{-\beta{\cal H}(\{\eta_r\})}$; $\beta=1/T$ is the inverse temperature, and we set $k_B=1$.
The thermal average of an operator ${\cal O}$ is given by
  \begin{align}
   \means{{\cal O}}=\frac{1}{Z}\sum_{\{\eta_r\}}{\rm Tr}_{\{c_j\}}\left[{\cal O}e^{-\beta {\cal H}}\right]=\means{\bar{\cal O}(\{\eta\})}_\eta,
  \end{align}
where we define $\bar{\cal O}(\{\eta_r\})=Z_f(\{\eta_r\})^{-1} {\rm Tr}_{\{c_j\}}[{\cal O} e^{-\beta {\cal H}(\{\eta_r\})}]$ and $\means{\cdots}_\eta=Z^{-1}\sum_{\{\eta_r\}}[\cdots]e^{-\beta F_f(\{\eta_r\})}$ with $Z$ being the partition function of the system.
In our calculations, we take the sum over configurations $\{\eta_r\}$ in the average $\means{\cdots}_\eta$ by performing Monte Carlo (MC) simulations so as to reproduce the distribution $e^{-\beta F_f(\{\eta_r\})}$.  
This admits the quantum MC (QMC) simulation which is free from the sign problem~\cite{PhysRevB.92.115122}.

In order to calculate the Raman spectrum at finite $T$, we employ the Loudon-Fleury (LF) approach~\cite{PhysRev.166.514,PhysRevLett.65.1068} by following previous $T=0$ studies~\cite{PhysRevLett.113.187201,PhysRevB.92.094439}:
the LF operator for the Kitaev model is given by
 ${\cal R}=\sum_{\gamma=x,y,z}\sum_{\means{jk}_\gamma}(\bm{\epsilon}_{\rm in}\cdot \bm{d}^\gamma)(\bm{\epsilon}_{\rm out}\cdot \bm{d}^\gamma)J^\gamma S_j^\gamma S_k^\gamma$,
where $\bm{\epsilon}_{\rm in}$ and $\bm{\epsilon}_{\rm out}$ are the polarization vectors of the incoming and outgoing photons and $\bm{d}^\gamma$ is the vector connecting sites on a NN $\gamma$ bond.
Using the LF operator, the Raman intensity is given by
$
I^{ll'}(\omega)=\frac{1}{N}\int_{-\infty}^{\infty}dt e^{i\omega t}\means{{\cal R}(t){\cal R}(0)},
$
where ${\cal R}(t)=e^{i{\cal H}t}{\cal R}e^{-i{\cal H}t}$ and $N$ is the number of sites; $l$ and $l'$ denote the directions of $\bm{\epsilon}_{\rm in}$ and $\bm{\epsilon}_{\rm out}$ in ${\cal R}$, respectively.
 Note that $I^{xx}=I^{yy}=I^{xy}\equiv I$ is satisfied in the isotropic case~\cite{PhysRevLett.113.187201}.
In terms of the Majorana fermions, the LF operator is described by a bilinear form of $c$ operators as
\begin{align}
 {\cal R}=\frac{1}{2}\sum_{jk}B_{jk}(\{\eta_r\}) c_j c_k,\label{eq:4}
\end{align}
  where $B(\{\eta_r\})$ is a Hermitian matrix with pure imaginary elements.
  Note that ${\cal R}(t)$ is simply given by $\frac{1}{2}\sum_{jk}B_{jk}(\{\eta_r\}) c_j(t) c_k(t)$ as all $\{\eta_r\}$ commute with the Hamiltonian.
  It is this property, which allows us to evaluate exactly the dynamical correlator of ${\cal R}$.
  Using Eq.~(\ref{eq:1}), we obtain 
\begin{align}
 {\cal R}=\frac{1}{2}\sum_{\lambda\lambda'} \left[
C_{\lambda\lambda'}\left(2f_\lambda^\dagger f_{\lambda'}-\delta_{\lambda\lambda'}\right)+D_{\lambda\lambda'}f_\lambda^\dagger f_{\lambda'}^\dagger + D_{\lambda'\lambda}^* f_\lambda f_{\lambda'}
 \right],\label{eq:2}
\end{align}
where $C_{\lambda\lambda'}=\sum_{jk} B_{jk} X_{j\lambda}^* X_{k\lambda'}$ and $D_{\lambda\lambda'}=\sum_{jk} B_{jk} X_{j\lambda}^* X_{k\lambda'}^*$.
  By applying Wick's theorem, we obtain the Raman intensity for a given configuration $\{\eta_r\}$ as 
  \begin{align}
 \bar{I}^{ll'}(\omega;\{\eta_r\})=&\frac{1}{N}\sum_{\lambda\lambda'}\Bigl[
 2\pi|C_{\lambda\lambda'}|^2 f(\varepsilon_\lambda)[1-f(\varepsilon_{\lambda'})]\delta(\omega+\varepsilon_\lambda-\varepsilon_{\lambda'})\nonumber\\
 &+\pi|D_{\lambda\lambda'}|^2[1-f(\varepsilon_\lambda)][1-f(\varepsilon_{\lambda'})]\delta(\omega-\varepsilon_\lambda-\varepsilon_{\lambda'}) \Bigr],\label{eq:6}
  \end{align}
where $\omega>0$.
  Finally, the thermal average is evaluated as $I^{ll'}(\omega)=\means{\bar{I}^{ll'}(\omega;\{\eta_r\})}_{\eta}$ using the QMC simulation.
  
  The  terms in Eq.~(\ref{eq:6}) describe two different Raman processes, which show different $T$ dependences via the Fermi distribution function $f(\varepsilon)$:
  the first term corresponds to the process (B) (Fig.~\ref{lattice}c) and the second term corresponds to the process (A) (Fig.~\ref{lattice}b).
  Thus, the $T$ dependence of the Raman intensity provides a good indicator of fermionic excitations in Kitaev QSLs.

Following our previous QMC study~\cite{PhysRevB.92.115122}, we have  performed more than 30000 MC steps for the measurements after 10000 MC steps for the thermalization using parallel tempering technique, for $N=2L^2$ clusters with $L=12$ and $20$.
The Raman intensity $I^{ll'}(\omega)$ is computed from 3000 samples during the 30000 MC steps.

{\bf Acknowledgments.}
  We thank M. Udagawa, K. Burch, P. Lemmens, B. Perreault, F. N. Burnell, N. B. Perkins, S. Kourtis, K. Ohgushi, and J. Yoshitake for fruitful discussions.
 J.K., D.K. and R.M. are very thankful to J.T. Chalker for collaborations on related work.
 We are especially grateful to L.~Sandilands and K.~Burch for sending us their experimental data on $\alpha$-RuCl$_3$.
 This work is supported by Grant-in-Aid for Scientific Research under Grant No. 24340076 and 15K13533, the Strategic Programs for Innovative Research (SPIRE), MEXT, the Computational Materials Science Initiative (CMSI), Japan, and the DFG via SFB 1143. The work of J.K. is supported by a Fellowship within the Postdoc-Program of the German Academic Exchange Service (DAAD). D.K. is supported by EPSRC Grant No. EP/M007928/1.
Parts of the numerical calculations are performed in the supercomputing systems in ISSP, the University of Tokyo.


\begin{thebibliography}{99} 
 \bibitem{Anderson1973153}
         Anderson, P. W.
         Resonating valence bonds: A new kind of insulator?
         \journal{Mater. Res. Bull.}{8}{153 -- 160}{1973}.
          
\bibitem{Kitaev2006}
         Kitaev, A.
         Anyons in an exactly solved model and beyond.
         \journal{Ann. Phys. (N. Y.)}{321}{2 -- 111}{2006}.
         
 \bibitem{Lacroix2011}
         Lacroix, C., Mendels, P. \& Mila, F.
         \textit{Introduction to Frustrated Magnetism.}
         Springer Series in Solid-State Sciences (Springer, Heidelberg, 2011).
         
\bibitem{PhysRevLett.114.147201}
        Sandilands, L.~J., Tian, Y., Plumb, K.~W., Kim, Y.-J. \& Burch, K.~S.
        Scattering continuum and possible fractionalized excitations in $\alpha$-${\mathrm{RuCl}}_{3}$.
         \journal{\PRL}{114}{147201}{2015}.


\bibitem{PhysRevLett.98.247201}
        Baskaran, G., Mandal, S. \& Shankar, R.
        Exact results for spin dynamics and fractionalization in the Kitaev model.
        \journal{\PRL}{98}{247201}{2007}.
        
\bibitem{PhysRevLett.112.207203}
        Knolle, J., Kovrizhin, D.L., Chalker, J.T. \& Moessner, R.
        Dynamics of a two-dimensional quantum spin liquid: Signatures of emergent Majorana fermions and fluxes.
        \journal{\PRL}{112}{207203}{2014}.


\bibitem{PhysRevLett.113.187201}
        Knolle, J., Chern, G.-W., Kovrizhin, D.~L., Moessner, R. \& Perkins, N.~B.
        Raman scattering signatures of Kitaev spin liquids in ${A}_{2}{\mathrm{IrO}}_{3}$ iridates with $A=\mathrm{Na}$ or Li.
        \journal{\PRL}{113}{187201}{2014}.
        
        
 \bibitem{PhysRevLett.113.197205}
         Nasu, J., Udagawa, M. \& Motome, Y.
         Vaporization of Kitaev Spin Liquids.
         \journal{\PRL}{113}{197205}{2014}.

 \bibitem{PhysRevB.92.115122}
         Nasu, J., Udagawa, M. \& Motome, Y.
         Thermal fractionalization of quantum spins in a Kitaev model: Temperature-linear specific heat and coherent transport of Majorana fermions.
         \journal{\PRB}{92}{115122}{2015}.

        
\bibitem{PhysRevLett.102.017205}
        Jackeli, G. \& Khaliullin, G.
        Mott insulators in the strong spin-orbit coupling limit: From Heisenberg to a quantum compass and Kitaev models.
         \journal{\PRL}{102}{017205}{2009}.

 \bibitem{PhysRevLett.108.127203}
         Singh, Y. \etal
         Relevance of the Heisenberg-Kitaev model for the honeycomb lattice iridates ${A}_{2}{\mathrm{IrO}}_{3}$.
         \journal{\PRL}{108}{127203}{2012}.
        
\bibitem{PhysRevLett.109.266406}
        Comin, R. \etal
        ${\mathrm{Na}}_{2}{\mathrm{IrO}}_{3}$ as a novel relativistic Mott insulator with a 340-meV gap.
         \journal{\PRL}{109}{266406}{2012}.
        
 \bibitem{PhysRevB.91.094422}
         Kubota, Y., Tanaka, H., Ono, T., Narumi, Y. \& Kindo, K.
         Successive magnetic phase transitions in $\alpha$-${\mathrm{RuCl}}_{3}$: XY-like frustrated magnet on the honeycomb lattice.
         \journal{\PRB}{91}{094422}{2015}.


\bibitem{PhysRevB.90.041112}
        Plumb, K.~W. \etal
        $\alpha$-${\mathrm{RuCl}}_{3}$: A spin-orbit assisted Mott insulator on a honeycomb lattice.
         \journal{\PRB}{90}{041112}{2014}.

\bibitem{Banerjee2015}
        Banerjee, A. \etal
        Proximate Kitaev quantum spin liquid behaviour in $\alpha$-${\mathrm{RuCl}}_{3}$.
         arXiv:1504.08037, \textit{unpublished}.

        
\bibitem{PhysRevLett.105.027204}
        Chaloupka, J., Jackeli, G. \& Khaliullin, G.
        Kitaev-Heisenberg model on a honeycomb lattice: Possible exotic phases in iridium oxides ${A}_{2}{\mathrm{IrO}}_{3}$.
         \journal{\PRL}{105}{027204}{2010}.

\bibitem{PhysRevB.84.100406}
        Reuther, J., Thomale, R. \& Trebst, S.
        Finite-temperature phase diagram of the Heisenberg-Kitaev model.
         \journal{\PRB}{84}{100406}{2011}.

        
\bibitem{PhysRevLett.110.097204}
        Chaloupka, J., Jackeli, G. \& Khaliullin, G.
        Zigzag magnetic order in the iridium oxide ${\mathrm{Na}}_{2}{\mathrm{IrO}}_{3}$.
         \journal{\PRL}{110}{097204}{2013}.

\bibitem{PhysRevLett.113.107201}
        Yamaji, Y., Nomura, Y., Kurita, M., Arita, R. \& Imada, M.
        First-principles study of the honeycomb-lattice iridates ${\mathrm{Na}}_{2}{\mathrm{IrO}}_{3}$ in the presence of strong spin-orbit interaction and electron correlations.
         \journal{\PRL}{113}{107201}{2014}.

\bibitem{1367-2630-16-1-013056}
        Katukuri, V.~M. \etal
        Kitaev interactions between $j=1/2$ moments in honeycomb ${\mathrm{Na}}_{2}{\mathrm{IrO}}_{3}$ are large and ferromagnetic: insights from \textit{ab initio} quantum chemistry calculations.
         \journal{New J. Phys.}{16}{013056}{2014}.

        
\bibitem{PhysRevB.88.035107}
        Foyevtsova, K., Jeschke, H.~O., Mazin, I.~I., Khomskii, D.~I. \& Valent\'{\i}, R.
        \textit{Ab initio} analysis of the tight-binding parameters and magnetic interactions in Na${}_{2}$IrO${}_{3}$.
         \journal{\PRB}{88}{035107}{2013}.

\bibitem{PhysRevB.91.241110}
        Kim, H.-S., Shankar, V.~V., Catuneanu, A. \& Kee, H.-Y.
        Kitaev magnetism in honeycomb ${\text{RuCl}}_{3}$ with intermediate spin-orbit coupling.
         \journal{\PRB}{91}{241110}{2015}.

\bibitem{PhysRevB.92.094439}
        Perreault, B., Knolle, J., Perkins, N.~B. \& Burnell, F.~J.
        Theory of raman response in three-dimensional Kitaev spin liquids: Application to $\ensuremath{\beta}$- and $\ensuremath{\gamma}$-${\mathrm{Li}}_{2}{\mathrm{IrO}}_{3}$ compounds.
         \journal{\PRB}{92}{094439}{2015}.
        
\bibitem{PhysRevB.57.8478}
        Sandvik, A.~W., Capponi, S., Poilblanc, D. \& Dagotto, E.
        Numerical calculations of the ${B}_{1g}$ Raman spectrum of the two-dimensional Heisenberg model.
         \journal{\PRB}{57}{8478--8493}{1998}.

 \bibitem{WenBook}
         Xiao-Gang Wen.
         Quantum Field Theory of Many-body Systems: From the Origin of Sound to an Origin of Light and Electrons. (see e.g. Chapter 8+9)
         {\it Oxford University Press} (2007).       

        
 \bibitem{PhysRevB.87.220407}
         Gretarsson, H. \etal
         Magnetic excitation spectrum of Na${}_{2}$IrO${}_{3}$ probed with resonant inelastic x-ray scattering.
         \journal{\PRB}{87}{220407}{2013}.


         
\bibitem{PhysRevB.76.193101}
        Chen, H.-D. \& Hu, J.
        Exact mapping between classical and topological orders in two-dimensional spin systems.
        \journal{\PRB}{76}{193101}{2007}.

\bibitem{PhysRevLett.98.087204}
        Feng, X.-Y., Zhang, G.-M. \& Xiang, T.
        Topological characterization of quantum phase transitions in a spin-$1/2$ model.
         \journal{\PRL}{98}{087204}{2007}.

\bibitem{PhysRev.166.514}
        Fleury, P.~A. \& Loudon, R.
        Scattering of light by one- and two-magnon excitations.
         \journal{\PR}{166}{514--530}{1968}.

\bibitem{PhysRevLett.65.1068}
        Shastry, B.~S. \& Shraiman, B.~I.
        Theory of Raman scattering in Mott-Hubbard systems.
         \journal{\PRL}{65}{1068--1071}{1990}.

\end{thebibliography}
\end{document}